\def\be{\begin{equation}}
\def\ee{\end{equation}}
\def\ba{\begin{array}}
\def\ea{\end{array}}

\documentclass[aps,amsmath,amssymb,amsfonts,showpacs,twocolumn,pra]{revtex4}
\usepackage{graphicx}
\usepackage{caption2}
\def\qed{\leavevmode\unskip\penalty9999 \hbox{}\nobreak\hfill
     \quad\hbox{\leavevmode  \hbox to.77778em{%
               \hfil\vrule   \vbox to.675em%
               {\hrule width.6em\vfil\hrule}\vrule\hfil}}
     \par\vskip3pt}

\begin{document}
\title{Unextendible maximally entangled bases and mutually unbiased bases}
\author{Bin Chen$^{1}$}
\author{Shao-Ming Fei$^{1,2}$}

\affiliation{$^1$School of Mathematical Sciences, Capital Normal University,
Beijing 100048, China\\
$^2$Max-Planck-Institute for Mathematics in the Sciences, 04103
Leipzig, Germany}

\begin{abstract}

We study unextendible maximally entangled basis in arbitrary bipartite spaces.
A systematic way of constructing a set of $d^{2}$
orthonormal maximally entangled states in
$\mathbb{C}^{d}\bigotimes\mathbb{C}^{d'}(\frac{d'}{2}<d<d')$ is provided.
The complementary space of the set of these $d^{2}$
orthonormal maximally entangled states contains no maximally entangled states that are
orthogonal to all of them. Furthermore,
we investigate mutually unbiased bases in which all the
bases are unextendible maximally entangled ones.
We present two unextendible maximally entangled bases in
$\mathbb{C}^{2}\bigotimes\mathbb{C}^{3}$ which are mutually
unbiased.

\end{abstract}

\pacs{ }
\maketitle

\section{Introduction}

Quantum entanglement plays vital role in quantum information
processing and has been extensively investigated in recent years
\cite{s1,g09}. It gives rise to the intrinsic distinction of quantum
mechanics, and is tightly related to some fundamental problems in quantum mechanics
such as reality and non-locality.
It has been shown that there could be
sets of product states which nevertheless display a form of
nonlocality \cite{s2,d03}.

Considerable theoretical results with useful applications have been obtained, ever since the
concept of unextendible product basis (UPB) in multipartite quantum
systems has been introduced in \cite{s2}. The UPB is a set of incomplete orthonormal
product basis whose complementary space has no product states.
It is shown that the mixed state on the subspace complementary
to a UPB is a bound entangled state. Moreover,
the states comprising a UPB are not distinguishable by local measurements and classical communication.

The UPB is generalized to unextendible maximally entangled basis (UMEB) \cite {s3}:
a set of orthonormal maximally entangled states in
$\mathbb{C}^{d}\bigotimes\mathbb{C}^{d}$ consisting of fewer than $d^2$ vectors which have no additional maximally
entangled vectors orthogonal to all of them.
It has been shown that there does not exist UMEBs for $d=2$. Explicit
examples are presented a 6-member UMEB for $d=3$ and a 12-member UMEB for $d=4$.

Another interestingly related problem is the mutually unbiased base (MUB).
Two orthonormal bases $\mathcal{B}_{1}=\{|b_{i}\rangle\}_{i=1}^{d}$ and
$\mathcal{B}_{2}=\{|c_{j}\rangle\}_{j=1}^{d}$ of $\mathbb{C}^{d}$
are said to be mutually unbiased (MU) if and only if
$$
|\langle b_{i}|c_{j}\rangle|=\frac{1}{\sqrt{d}},~~~\forall\, i,j=1,2,\cdots,d.
$$
If a physical system is prepared in an eigenstate of basis $\mathcal{B}_{1}$
and measured in basis $\mathcal{B}_{2}$, then all outcomes are equally probable.
A set of orthonormal bases
$\{\mathcal{B}_{1}, \mathcal{B}_{2},...,\mathcal{B}_{m}\}$ in $C^d$
is called a set of mutually unbiased
bases (MUBs) if every pair of bases in the set is mutually unbiased.

MUBs have useful applications in many quantum information processing,
such as quantum state tomography \cite{s5,qst1,qst2}, cryptographic
protocols \cite{cp1,cp2}, and the mean king¡¯s problem \cite{mk}.
The maximum number $N(d)$ of MUBs is no more than $d+1$.
It has been shown that $N(d)=d+1$ when $d$ is a prime power
\cite{s5}. However, when $d$ is a composite number, $N(d)$ is still
unknown. Even for $d=6$, we do not know whether there exist four MUBs
or not \cite{s6,s7,s8,s9}. Hence
the research on $N(6)$ and construction of MUBs in $\mathbb{C}^{6}$ is
of great importance.

In this paper, we first study UMEB in arbitrary
bipartite spaces $\mathbb{C}^{d}\bigotimes\mathbb{C}^{d'}$.
We provide a systematic way in constructing
$d^{2}$-member UMEBs in $\mathbb{C}^{d}\bigotimes\mathbb{C}^{d'}$ ($\frac{d'}{2}<d<d'$).
We show that the subspace complementary to the
$d^{2}$-member UMEB contains no states of Schmidt rank higher than
$d-1$. Thus we answer the question proposed by S. Bravyi and J. A.
Smolin in \cite{s3} whether there exist any ``nonsquare" UMEBs, i.e. UMEBs
in the bipartite spaces with different dimensions.
Moreover, we construct two
complete UMEBs in $\mathbb{C}^{2}\bigotimes\mathbb{C}^{3}$ which are
mutually unbiased.

\section{UMEBs in $\mathbb{C}^{d}\bigotimes\mathbb{C}^{d'}\,(\frac{d'}{2}<d<d')$}

We first study unextendible maximally entangled basis in bipartite spaces with different dimensions.

\textbf{Definition.} A set of states
\{$|\phi_{i}\rangle\in\mathbb{C}^{d}\bigotimes\mathbb{C}^{d'}:i=1,2,\cdots,n,\,n<dd'$\}
is called an $n$-number UMEB if and only if

(i) $|\phi_{i}\rangle$, $i=1,2,\cdots,n$, are maximally entangled;

(ii) $\langle\phi_{i}|\phi_{j}\rangle=\delta_{ij}$;

(iii) if $\langle\phi_{i}|\psi\rangle=0$ for all $i=1,2,\cdots,n$, then $|\psi\rangle$ cannot be maximally entangled.

Here state $|\psi\rangle$ is said to be a $d\otimes d'$ maximally entangled state if and only if
for arbitrary given orthonormal complete basis \{$|i_{A}\rangle$\}
of subsystem A, there exists an orthonormal basis
\{$|i_{B}\rangle$\} of subsystem B such that $|\psi\rangle$ can be written as
$|\psi\rangle=\frac{1}{\sqrt{d}}\sum_{i=0}^{d-1}|i_{A}\rangle\otimes|i_{B}\rangle$ \cite{s4}.

We first construct UMEB in $\mathbb{C}^{2}\bigotimes\mathbb{C}^{3}$, then generalize the
construction to the case in $\mathbb{C}^{d}\bigotimes\mathbb{C}^{d'}(\frac{d'}{2}<d<d')$.
Consider the following four maximally entangled states in
$\mathbb{C}^{2}\bigotimes\mathbb{C}^{3}$:
\begin{equation}\label{2x3}
\begin{split}
|\phi_{0}\rangle &=\frac{1}{\sqrt{2}}(|0\rangle|0'\rangle+|1\rangle|1'\rangle),\\
|\phi_{i}\rangle &=(\sigma_{i}\otimes I_{3})|\phi_{0}\rangle,~~ i=1,2,3,
\end{split}
\end{equation}
where $\{|0\rangle,|1\rangle\},\{|0'\rangle,|1'\rangle,|2'\rangle\}$
are the orthonormal bases in $\mathbb{C}^{2}$ and $\mathbb{C}^{3}$
respectively, $\sigma_{i}$, $i=1,2,3$, are the Pauli matrices. It can be
easily checked that the above four states are mutually orthogonal.
Now we prove that if there exits $|\psi\rangle$ such that
$\langle\phi_{i}|\psi\rangle=0$, $\forall ~i=0,1,2,3$, then $|\psi\rangle$ must
be a product state.

$|\psi\rangle$ can be generally written in the Schmidt decomposition form,
\begin{equation}
|\psi\rangle=(U\otimes V)(\sqrt{\lambda_{0}}|0\rangle|0'\rangle+\sqrt{\lambda_{1}}|1\rangle|1'\rangle),
\end{equation}
where $\lambda_{0}>0$, $\lambda_{1}>0$, $\lambda_{0}+\lambda_{1}=1$,
$U$ and $V$ are unitary operators, $U=(u_{ij})_{2\times2}$, $V=(v_{ij})_{3\times3}$
in the ordered bases $\{|0\rangle,|1\rangle\}$ and
$\{|0'\rangle,|1'\rangle,|2'\rangle\}$ respectively.

The orthogonal relations between $|\psi\rangle$ and $|\phi_{i}\rangle$
give rise to the following equations:
$$
\ba{l}
\sqrt{\lambda_{0}}\langle0|\sigma_{i}U|0\rangle\langle0'|V|0'\rangle+
\sqrt{\lambda_{1}}\langle0|\sigma_{i}U|1\rangle\langle0'|V|1'\rangle \\[2mm]
+\sqrt{\lambda_{0}}\langle1|\sigma_{i}U|0\rangle\langle1'|V|0'\rangle+
\sqrt{\lambda_{1}}\langle1|\sigma_{i}U|1\rangle\langle1'|V|1'\rangle=0,
\ea
$$
$i=0,1,2,3,$ where $\sigma_{0}=I_{2}$. Hence we have
\begin{eqnarray*}
\sqrt{\lambda_{0}}u_{11}v_{11}+\sqrt{\lambda_{1}}u_{12}v_{12}+
\sqrt{\lambda_{0}}u_{21}v_{21}+\sqrt{\lambda_{1}}u_{22}v_{22}=0,\\
\sqrt{\lambda_{0}}u_{21}v_{11}+\sqrt{\lambda_{1}}u_{22}v_{12}+
\sqrt{\lambda_{0}}u_{11}v_{21}+\sqrt{\lambda_{1}}u_{12}v_{22}=0,\\
\sqrt{\lambda_{0}}u_{11}v_{11}+\sqrt{\lambda_{1}}u_{12}v_{12}-
\sqrt{\lambda_{0}}u_{21}v_{21}-\sqrt{\lambda_{1}}u_{22}v_{22}=0,\\
\sqrt{\lambda_{0}}u_{21}v_{11}+\sqrt{\lambda_{1}}u_{22}v_{12}-
\sqrt{\lambda_{0}}u_{11}v_{21}-\sqrt{\lambda_{1}}u_{12}v_{22}=0.
\end{eqnarray*}
The set of above equations can be expressed as
$M\overrightarrow{v}=\overrightarrow{0}$, where
\begin{equation*}
M=\begin{pmatrix}
U & AU\\U & -AU
\end{pmatrix}
\begin{pmatrix}
W &  \\  & W
\end{pmatrix},
\end{equation*}
\begin{equation*}
A=\begin{pmatrix}
0&1\\1&0
\end{pmatrix},~~~
W=\begin{pmatrix}
\sqrt{\lambda_{0}}&\\&\sqrt{\lambda_{1}}
\end{pmatrix},
\end{equation*}
and
$$
\overrightarrow{v}=(v_{11},v_{12},v_{21},v_{22})^{T},
$$
where $T$ denotes matrix transpose.

Since $det M\neq0$, we obtain
$\overrightarrow{v}=\overrightarrow{0}$, which implies that $det\,V=0$, and hence
$V$ cannot be unitary. Thus (\ref{2x3}) is a 4-member UMEB in
$\mathbb{C}^{2}\bigotimes\mathbb{C}^{3}$, whose complementary space
contains no entangled states.

We now construct UMEBs in
$\mathbb{C}^{d}\bigotimes\mathbb{C}^{d'}$, $\frac{d'}{2}<d<d'$, by
using the approach for the case of
$\mathbb{C}^{2}\bigotimes\mathbb{C}^{3}$. Let
$\{|i\rangle\}_{i=0}^{d-1}$ and $\{|i'\rangle\}_{i=0}^{d'-1}$ be
orthonormal bases of $\mathbb{C}^{d}$ and $\mathbb{C}^{d'}$
respectively. Consider a set of unitary operators, which forms a
basis of the operator space on $\mathbb{C}^{d}$:
\begin{equation}
U_{nm}=\sum_{k=0}^{d-1}\zeta_{d}^{nk}|k\oplus m\rangle\langle k|,~~ n,m=0,1,\cdots,d-1,
\end{equation}
where $\zeta_{d}=e^{\frac{2\pi\sqrt{-1}}{d}}$, $k\oplus m$ denotes
$(k+m)$ mod $d$. These operators satisfy
$$
Tr(U_{n'm'}^{\dag}U_{nm})=d\,\delta_{n'n}\delta_{m'm}.
$$
Applying this set of unitary operators to the first party of a given
maximally entangled state in
$\mathbb{C}^{d}\bigotimes\mathbb{C}^{d'}$, we get $d^{2}$ mutually
orthonormal maximally entangled states:
\begin{equation}\label{mn}
|\Phi_{nm}\rangle=(U_{nm}\otimes I_{d'})|\Phi\rangle,~~ n,m=0,1,\cdots,d-1,
\end{equation}
where
$|\Phi\rangle=\frac{1}{\sqrt{d}}\sum_{p=0}^{d-1}|p\rangle|p'\rangle$.
We now prove that if there  exists $|\Psi\rangle$ in
$\mathbb{C}^{d}\bigotimes\mathbb{C}^{d'}$, $\frac{d'}{2}<d<d'$, such
that $\langle\Phi_{nm}|\Psi\rangle=0$ for all $n, m$, then
$|\Psi\rangle$ cannot be of Schmidt rank $d$, i.e. $|\Psi\rangle$ must not
be maximally entangled.

$|\Psi\rangle$ can generally written as
$$
|\Psi\rangle=(U\otimes
V)\sum_{p=0}^{d-1}\sqrt{\lambda_{p}}|p\rangle|p'\rangle,
$$
where $\lambda_{p}>0$ for all $p$, $\sum_{p=0}^{d-1}\lambda_{p}=1$,
$U=(u_{ij})_{d\times d}$ and $V=(v_{ij})_{d'\times d'}$ are unitary
matrices in the given ordered bases respectively. If
$\langle\Phi_{nm}|\Psi\rangle=0$ for all $n$ and $m$, one gets $d^{2}$ equations:
\begin{equation}
\begin{split}
\langle\Phi_{nm}|\Psi\rangle&=\frac{1}{\sqrt{d}}\sum_{k,p}\zeta_{d}^{-nk}\sqrt{\lambda_{p}}\langle k\oplus m|U|p\rangle\langle k'|V|p'\rangle\\
&=0,~~~ n,m=0,1,\cdots,d-1.
\end{split}
\end{equation}
These equations can be expressed in a compact form: $M\overrightarrow{v}=\overrightarrow{0}$, where
\begin{widetext}
\begin{equation*}
M=\begin{pmatrix}
U  &   AU &  A^{2}U  &  \cdots &  A^{d-1}U\\
U &   \zeta_{d}^{-1}AU &  \zeta_{d}^{-2}A^{2}U  &  \cdots &  \zeta_{d}^{-(d-1)}A^{d-1}U\\
U  &   \zeta_{d}^{-2}AU &  \zeta_{d}^{-4}A^{2}U  &  \cdots &  \zeta_{d}^{-2(d-1)}A^{d-1}U\\
\vdots & \vdots & \vdots & & \vdots\\
U  &   \zeta_{d}^{-(d-1)}AU &  \zeta_{d}^{-2(d-1)}A^{2}U  &  \cdots &  \zeta_{d}^{-(d-1)^{2}}A^{d-1}U
\end{pmatrix}
\begin{pmatrix}
W& & & & \\
 &W& & & \\
 & &W& & \\
 & & &\ddots& \\
 & & & &W
\end{pmatrix},
\end{equation*}
\end{widetext}
$A$ is a permutation matrix,
\begin{equation*}
A=\begin{pmatrix}
0&1&0&\cdots &0\\
0&0&1&\cdots &0\\
\vdots & \vdots & \vdots & &\vdots\\
0&0&0&\cdots &1\\
1&0&0&\cdots &0
\end{pmatrix},
\end{equation*}
$W=diag\{\sqrt{\lambda_{0}},\sqrt{\lambda_{1}},\cdots
\sqrt{\lambda_{d-1}}\}$,
$\overrightarrow{v}=(v_{11},v_{12},\cdots,v_{1d},v_{21},v_{22},\cdots,v_{2d},\cdots,v_{d1},v_{d2},\cdots,v_{dd})^{T}$.
Since $M$ can further written as
\begin{widetext}
\begin{equation*}
M=\begin{pmatrix}
I  &   I &  I  &  \cdots &  I\\
I &   \zeta_{d}^{-1}I &  \zeta_{d}^{-2}I  &  \cdots &  \zeta_{d}^{-(d-1)}I\\
I  &   \zeta_{d}^{-2}I &  \zeta_{d}^{-4}I  &  \cdots &  \zeta_{d}^{-2(d-1)}I\\
\vdots & \vdots & \vdots & & \vdots\\
I  &   \zeta_{d}^{-(d-1)}I &  \zeta_{d}^{-2(d-1)}I &  \cdots &  \zeta_{d}^{-(d-1)^{2}}I
\end{pmatrix}
\begin{pmatrix}
I& & & & \\
 &A& & & \\
 & &A^{2}& & \\
 & & &\ddots& \\
 & & & &A^{d-1}
\end{pmatrix}
\begin{pmatrix}
U& & & & \\
 &U& & & \\
 & &U& & \\
 & & &\ddots& \\
 & & & &U
\end{pmatrix}
\begin{pmatrix}
W& & & & \\
 &W& & & \\
 & &W& & \\
 & & &\ddots& \\
 & & & &W
\end{pmatrix},\
\end{equation*}
\end{widetext}
it is easily verified that $detM\neq0$, which implies that
$\overrightarrow{v}=\overrightarrow{0}$. Due to that $d'<2d$, we have
$det\,V=0$ and $V$ cannot be unitary.
Therfore the $d^{2}$ mutually
orthonormal maximally entangled states (\ref{mn})
constitute a $d^{2}$-member UMEB in
$\mathbb{C}^{d}\bigotimes\mathbb{C}^{d'}$ for $\frac{d'}{2}<d<d'$.

Like the UMEBs in $\mathbb{C}^{d}\bigotimes\mathbb{C}^{d}$, our
``nonsquare" UMEBs have also interesting and useful
applications in quantum information processing. For example, as it has
been noted in \cite{s3}, the relation of the complementary
space to unital channels would be broken. Let
$\{|\Phi_{i}\rangle:~i=1,2,\cdots,d^{2}\}$ be the set of $d^{2}$-member
UMEB constructed above in $\mathbb{C}^{d}\bigotimes\mathbb{C}^{d'}$, $\frac{d'}{2}<d<d'$.
The state
\begin{equation}\label{6}
\rho^{\perp}=\frac{1}{dd'-d^{2}}(I-\sum_{i=1}^{d^{2}}|\Phi_{i}\rangle\langle\Phi_{i}|),
\end{equation}
on a $d\otimes d'$ bipartite system
$\mathcal{H}_{A}\bigotimes\mathcal{H}_{B}$ corresponds to a
completely positive  map
$\Lambda:\,\mathrm{B}(\mathcal{H}_{A})\longrightarrow\mathrm{B}(\mathcal{H}_{B})$
by Jamiolkowski isomorphism \cite{s4}, and $\Lambda$ is a quantum channel
since $Tr_{B}(\rho^{\perp})=\frac{1}{d}I_{d}$. But the corresponding
channel would not be unital since
$$
Tr_{A}(\rho^{\perp})=\frac{1}{d'-d}(I_{d'}-\sum_{i=0}^{d-1}|i'\rangle\langle
i'|)\neq\frac{1}{d'}I_{d'}.
$$
Besides, as $\rho^{\perp}$
cannot be a mixture of maximally entangled states, $\Lambda$ cannot
be convex mixtures of unitary operators.

\section{MUBs from UMEBs in $\mathbb{C}^{2}\bigotimes\mathbb{C}^{3}$}

MUBs, UPBs and UMEBs have significant applications in quantum information
processing. In this section we study the possibility that all the bases in MUBs
are also UMEBs. We construct two UMEBs in
$\mathbb{C}^{2}\bigotimes\mathbb{C}^{3}$ which are mutually
unbiased.

By using (\ref{2x3}) we have the first UMEB in $\mathbb{C}^{2}\bigotimes\mathbb{C}^{3}$:
\begin{equation}\label{b1}
\begin{split}
|\phi_{0}\rangle &=\frac{1}{\sqrt{2}}(|0\rangle|0'\rangle+|1\rangle|1'\rangle),\\
|\phi_{i}\rangle &=(\sigma_{i}\otimes I_{3})|\phi_{0}\rangle,~~ i=1,2,3,\\
|\phi_{4}\rangle &=(\frac{1}{2}|0\rangle+\frac{\sqrt{3}}{2}|1\rangle)\otimes|2'\rangle,\\
|\phi_{5}\rangle &=(\frac{\sqrt{3}}{2}|0\rangle-\frac{1}{2}|1\rangle)\otimes|2'\rangle.
\end{split}
\end{equation}

Starting with another basis in $\mathbb{C}^{3}$:
\begin{equation*}
\begin{split}
|x'\rangle &=\frac{1}{\sqrt{3}}(|0'\rangle+\frac{1+\sqrt{3}i}{2}|1'\rangle+|2'\rangle),\\
|y'\rangle &=\frac{1}{\sqrt{3}}(\frac{-\sqrt{3}+i}{2}|0'\rangle+i|1'\rangle-i|2'\rangle),\\
|z'\rangle &=\frac{1}{\sqrt{3}}(-|0'\rangle+|1'\rangle+\frac{1+\sqrt{3}i}{2}|2'\rangle),
\end{split}
\end{equation*}
where $i=\sqrt{-1}$, we can construct another UMEB in $\mathbb{C}^{2}\bigotimes\mathbb{C}^{3}$,
\begin{equation}\label{b2}
\begin{split}
|\psi_{j}\rangle &=\frac{1}{\sqrt{2}}(\sigma_{j}\otimes I_{3})(|0\rangle|x'\rangle+|1\rangle|y'\rangle),~~j=0,1,2,3,\\
|\psi_{4}\rangle &=\frac{1}{\sqrt{2}}(\frac{1+\sqrt{3}i}{2}|0\rangle+\frac{\sqrt{3}-i}{2}|1\rangle)\otimes|z'\rangle,\\
|\psi_{5}\rangle &=\frac{1}{\sqrt{2}}(\frac{\sqrt{3}-i}{2}|0\rangle+\frac{1+\sqrt{3}i}{2}|1\rangle)\otimes|z'\rangle.
\end{split}
\end{equation}
It is directly verified that the above two bases (\ref{b1}) and (\ref{b2}) are mutually unbiased.

\section{Conclusion and discussion}

We have studied UMEB in arbitrary bipartite spaces, and
provided explicit construction of UMEBs in
$\mathbb{C}^{d}\bigotimes\mathbb{C}^{d'}({d'}/{2}<d<d')$. We are not sure
if there exist UMEBs in
$\mathbb{C}^{d}\bigotimes\mathbb{C}^{d'}$ for $d\leq {d'}/{2}$,
and whether in this case the set of $d^{2}$-member
orthonormal maximally entangled basis we constructed is unextendible
or not. If they are unextendible, one gets that the
entanglement of assistance (EoA) \cite{eoa} is strictly smaller than
the asymptotic EoA \cite{aeoa} as the UMEBs do in
$\mathbb{C}^{d}\bigotimes\mathbb{C}^{d}$ \cite{s3}. In fact, the
asymptotic EoA of $\rho^{\perp}$ in (\ref{6}) is equal to $\log d$ when
$d\leq {d'}/{2}$, since $S(Tr_{A}\ \rho^{\perp})=\mathrm{log}
(d'-d)\geq S(Tr_{B}\ \rho^{\perp})=\mathrm{log}\ d$, and the EoA of
$\rho^{\perp}$ is strictly smaller than $\log d$.

Moreover, we have presented two UMEBs in
$\mathbb{C}^{2}\bigotimes\mathbb{C}^{3}$ which are mutually
unbiased. It would be interesting to investigate the implications
and applications in quantum state tomography and cryptographic
protocols for such MUBs in which all the bases are UMEBs.

\bigskip
\noindent{\bf Acknowledgments}\, \,
This work is supported by the NSFC under number 11275131.

\end{document}